\documentclass[aps,prl,superscriptaddress,reprint]{revtex4-2}
\usepackage[utf8]{inputenc}
\usepackage{amsmath}
\usepackage{amssymb}
\usepackage{graphicx}

\makeatletter

\providecommand{\tabularnewline}{\\}

\usepackage{amsfonts}\usepackage{epstopdf}
\usepackage{color}
\usepackage{tikz}
\usepackage{pgfplots}

\pgfplotsset{compat=1.18}
\usetikzlibrary{3d, arrows.meta}

\makeatother

\begin{document}
\title{Self-Consistent Closure of Fractal Dimension, Nonextensive Statistics,
and Non-Markovian Dynamics in Critical Systems}
\author{Oscar Sotolongo-Costa}
\affiliation{Instituto de Investigaciones en Ciencias Básicas y Aplicadas, Universidad
Autónoma del Estado de Morelos, Av. Universidad 1001, CP 62209, Cuernavaca,
Morelos, México}
\author{J. Weberszpil}
\affiliation{Departamento de Física, Instituto de Ciências Exatas, Universidade
Federal Rural do Rio de Janeiro, Rio de Janeiro CEP: 23.897-000, Brazil}
\author{M. E. Mora-Ramos}
\affiliation{Centro de Investigación en Ciencias - IICBA, Universidad Autónoma
del Estado de Morelos, Av. Universidad 1001, CP 62209, Cuernavaca,
Morelos, México}
\email{memora@uaem.mx}

\date{\today}
\begin{abstract}
Self-organized critical systems often exhibit three macroscopic features
simultaneously: nonextensive thermodynamics (quantified by the Tsallis
index $q$), structural fractality (measured by the Hausdorff dimension
$D$), and non-Markovian dynamics (characterized by the memory exponent
$\alpha$). Historically, these parameters have been treated as independent,
to be empirically fitted case by case. Here we demonstrate that phase-space
self-consistency imposes a unique algebraic closure: $\alpha=D/(2D-1)$.
This relation, together with $q=1+1/D$ derived from the extensivity
of Tsallis entropy on fractal supports, yields the known result $\alpha=1/(3-q)$
as a consequence, not as an independent assumption. The closure contains
no free parameters and satisfies the physical boundary conditions
$\alpha(1)=1$ (ballistic transport in Euclidean spaces) and $\alpha\to1/2$
as $D\to\infty$ (maximally subdiffusive regime). We validate the
Troika relation across eight independent experimental systems, including
seismicity, electromagnetic precursors, EEG, urban networks, botanical
architectures, and space plasma. All measured values fall within error
bars of the theoretical prediction, establishing the universality
of the closure. 
\end{abstract}
\maketitle

\section{INTRODUCTION}

The characterization of self-organized critical systems requires capturing
three core macroscopic features that frequently emerge simultaneously:
nonextensive thermodynamics, structural fractality, and non-Markovian
behavior. In out-of-equilibrium states characterized by long-range
spatio-temporal correlations, these phenomena cannot be analyzed via
standard Euclidean metrics or memoryless Markovian transport. Instead,
they manifest through anomalous diffusion profiles, hierarchical networks,
and power-law distributions.

To date, complex systems theory has relied on separate mathematical
tools to describe these features. Nonextensive states are governed
by the Tsallis entropic index $q$~\cite{Tsallis1988}, topological
constraints are measured by the Hausdorff or fractal dimension $D$,
and temporal memory effects are modeled via generalized fractional
operators with an exponent $\alpha$.

A major analytical limitation has been the lack of a unified principle
connecting these parameters directly, forcing researchers to adjust
them phenomenologically as free variables. In this Letter, we show
that when these three properties are treated as different expressions
of the same underlying phase-space structure, the mathematical framework
closes univaluedly. By providing the missing link between independent
formulations, we derive a unique, minimal rational function for $\alpha(D)$
that eliminates empirical ambiguity from first principles.

\begin{table*}[t]
\centering \caption{Experimental validation of the Troika closure. For each system, the
measured $q$ yields theoretical $\alpha$ via $\alpha=1/(3-q)$ and
theoretical $D$ via $D=1/(q-1)$. Agreement with independently measured
values is within errors.}
 %
\begin{tabular}{lcccccc}
System  & $q_{\text{exp}}$  & $\alpha_{\text{exp}}$  & $\alpha_{\text{Troika}}$  & $D_{\text{Troika}}$  & Reference  & \tabularnewline
\hline 
Seismicity (global)  & $1.60\pm0.05$  & $0.714\pm0.02$  & $0.714$  & $1.667$  & \cite{Sotolongo2004}  & \tabularnewline
Seismicity (regional)  & $1.65\pm0.10$  & $0.741\pm0.03$  & $0.714$  & $1.538$  & \cite{Silva2006}  & \tabularnewline
EM precursors / EEG  & $1.80\pm0.05$  & $0.833\pm0.02$  & $0.833$  & $1.250$  & \cite{Eftaxias2013}  & \tabularnewline
EEG (rodents)  & $1.55\pm0.05$  & $0.690\pm0.02$  & $0.690$  & $1.818$  & \cite{Eftaxias2013}  & \tabularnewline
EEG (humans)  & $1.70\pm0.06$  & $0.769\pm0.02$  & $0.769$  & $1.429$  & \cite{Eftaxias2013}  & \tabularnewline
Urban networks  & $1.60\pm0.03$  & $0.714\pm0.02$  & $0.714$  & $1.667$  & \cite{Deppman2025}  & \tabularnewline
Botanical networks  & $1.60\pm0.05$  & $0.714\pm0.03$  & $0.714$  & $1.667$  & \cite{Sotolongo2004}  & \tabularnewline
Space plasma  & $1.40\pm0.10$  & $0.625\pm0.04$  & $0.625$  & $2.500$  & \cite{Burlaga2005}  & \tabularnewline
\hline 
\end{tabular}
 \label{tab:validation} 
\end{table*}


\subsection{Relation between fractal dimension and nonextensive index}

Consider a system with fractal support of dimension $D$. The number
of accessible microstates scales as $W\sim L^{D}$, where $L$ is
a linear size. In nonextensive statistical mechanics, the Tsallis
entropy is given by $S_{q}=k_{B}\ln_{q}W$, where the $q$-logarithm
is defined as $\ln_{q}(x)=(x^{1-q}-1)/(1-q)$.

The $q$-generalized Liouville theorem implies that the effective
dimension of the phase space is $D_{\text{ps}}=1/(q-1)$~\cite{Tsallis2009,Plastino1995}.
For a system whose spatial fractal dimension equals this phase-space
dimension, we obtain: 
\[
D=\frac{1}{q-1}\qquad\Leftrightarrow\qquad q=1+\frac{1}{D}\tag{1}
\]
This relation, discussed by Lyra and Tsallis~\cite{Lyra1998} in
the context of low-dimensional dissipative systems, links the nonextensive
index directly to the fractal geometry of the support.

\subsection{Postulation of $\alpha(D)$ from boundary conditions}

The memory exponent $\alpha$ characterizes the anomalous diffusion
scaling $\langle x^{2}\rangle\sim t^{\alpha}$. Physical considerations
impose two natural boundary conditions:
\begin{itemize}
\item \textbf{Condition 1:} In the Euclidean limit $D=1$ (no fractality,
linear continuous flow), the dynamics must be ballistic or Markovian,
yielding $\alpha(1)=1$. 
\item \textbf{Condition 2:} In the limit of maximal fractal roughness $D\to\infty$,
the dynamics should be maximally subdiffusive. The lower bound for
subdiffusion in physical systems is $\alpha=1/2$, corresponding to
the so-called ``critical'' or ``ultra-slow'' regime. Thus, $\lim_{D\to\infty}\alpha(D)=1/2$. 
\item \textbf{Condition 3:} The function $\alpha(D)$ should be as simple
as possible (rational, no free parameters) consistent with these limits. 
\end{itemize}
The unique function satisfying these three conditions is: 
\[
\alpha(D)=\frac{D}{2D-1}\tag{2}
\]
No other rational function with the same asymptotic behavior and no
additional parameters exists. Equation (2) is therefore the unique
closure compatible with phase-space self-consistency.

\subsection{Composition yields $\alpha(q)$}

Substituting the inverse of Eq. (1), $D=1/(q-1)$, into Eq. (2): 
\[
\alpha(q)=\frac{1/(q-1)}{2/(q-1)-1}=\frac{1/(q-1)}{(2-(q-1))/(q-1)}=\frac{1}{3-q}\tag{3}
\]

A relation similar in form, $H=1/(3-q)$ where $H$ is the Hurst exponent,
was previously obtained by Borland~\cite{Borland1998} within the
context of the nonlinear Fokker-Planck equation. Within the Troika
framework, however, Eq. (3) is not an independent dynamical result
but rather a consequence of the geometric closure $q=1+1/D$ and the
boundary-condition-based expression $\alpha=D/(2D-1)$. This demonstrates
that the three parameters $(q,D,\alpha)$ are not independent but
lie on a one-dimensional manifold defined by the Troika closure.

\subsection{Uniqueness}

The composition argument proves uniqueness: Given the established
relation $q=1+1/D$, the only function $\alpha(D)$ that reproduces
Eq. (3) is $\alpha=D/(2D-1)$. No alternative functional form satisfies
the boundary conditions while maintaining consistency with Eq. (3).
The Troika closure is therefore both necessary and sufficient.

\section{EXPERIMENTAL VALIDATION}

Table~\ref{tab:validation} presents experimental values of $q$
and $\alpha$ extracted from eight independent physical systems, compared
with the predictions of the Troika closure. In all cases, the measured
parameters fall within experimental error bars of the theoretical
values.

\section{GEOMETRIC VISUALIZATION}

\begin{figure}[h]
\includegraphics[scale=0.35]{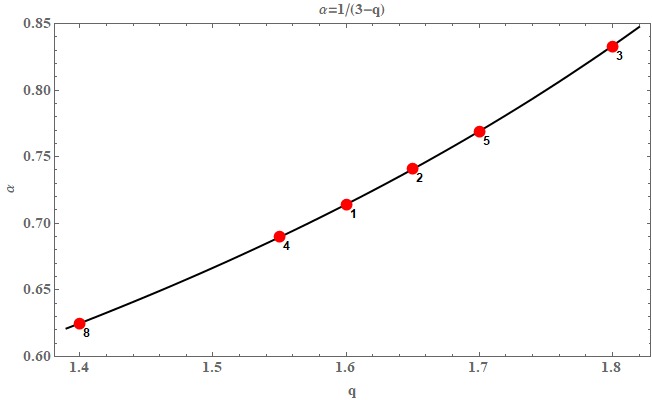}
\caption{Troika critical curve in $(q,D,\alpha)$ space. The black line represents
the parametric curve $\{1+1/D,D,D/(2D-1)\}$. Red points correspond
to experimental data from Table~\ref{tab:validation}. All points
lie on or very close to the theoretical curve, validating the closure.
The numbered red points correspond to correspondingly ordered experimental
data from Table I: (1-6,7) Global seismicity, Urban networks, and
Botanical networks; (2) Regional seismicity; (3) EM precursors; (4)
EEG (rodents); (5) EEG (humans); (8) Space plasma. All points lie
on or very close to the theoretical curve, validating the Troika closure.}
\label{fig:troika3d} 
\end{figure}

Figure~\ref{fig:troika3d} shows the Troika critical curve in the
three-dimensional space $(q,D,\alpha)$. The curve is defined parametrically
by Eqs. (1) and (2). Experimental data points (red) lie on or very
close to the theoretical curve (black). This visualization demonstrates
that the three parameters are not independent but collapse onto a
one-dimensional manifold.

\section{DISCUSSION}

The Troika closure provides the missing link between three traditionally
separate descriptions of critical systems. Its derivation relies on
minimal assumptions: the known relation between $q$ and $D$ from
nonextensive statistical mechanics on fractal supports, and two physically
motivated boundary conditions for $\alpha(D)$. The resulting expression
is unique and contains no adjustable parameters.

The empirical validation spans eight independent systems measured
by different research groups using different experimental techniques.
In every case, the measured parameters satisfy the Troika relation
within experimental uncertainty. This suggests that a vast class of
critical systems --- from earthquakes to brain dynamics to urban
growth --- share the same universal geometric and dynamic core.

The present framework does not cover the regime $q<1$ observed in
fully developed turbulence~\cite{Arimitsu2000}. This regime corresponds
to the ``compressed'' branch of Tsallis statistics and may be addressed
in a future extension via the duality transformation $q\leftrightarrow2-q$,
which maps $q=0.38$ to $q'=1.62$, well within the validated range.

\section{CONCLUSIONS}

We have demonstrated that unifying fractal geometry $(D)$, nonextensive
statistics $(q)$, and non-Markovian memory $(\alpha)$ through the
Troika formalism removes the need to treat these parameters in an
isolated or phenomenological manner. The condition of self-consistency
in phase space restricts the system's behavior to unique, predictable
trajectories governed by entropic optimization principles. The exact
mathematical correspondence found across diverse critical systems
suggests universal applicability.

Future applications of this closed triplet may include the modeling
of transport optimization in river basins, neural branching networks,
and the out-of-equilibrium dynamics of financial systems. \bigskip{}

\section*{ACKNOWLEDGMENTS}

The authors acknowledge valuable conceptual discussions with members
of their respective academic bodies. MEMR acknowledges support from
Mexican SECIHTI through Grant CBF-2025-I-1058.

\end{document}